# Josephson dynamics of a Bose-Einstein condensate in an accelerated double well potential

Aranya B. Bhattacherjee
Department of Physics, Atma Ram Sanatan Dharma College
(University of Delhi, South Campus),
Dhaula Kuan, New Delhi- 110021
India,
Email: abhattac@bol.net.in

**Abstract**: Motivated by recent experiment on Bloch Oscillation of Bose-Einstein condensates (BEC) in accelerated optical lattices, we consider the Josephson dynamics of a BEC in an accelerated double well potential. We show that acceleration suppresses coherent population/phase oscillation between the two wells. Accelerating the double well renders the Josephson coupling energy $E_J$ time dependent and this emerges as a source of dissipation. This dissipative mechanism helps to stabilize the system. The results are used to interpret a recent experimental result [reference 16].

PACS No: 03.75.Lm

## Introduction

The Josephson effect is an important quantum phenomenon, first discovered in superconductors and later in superfluid helium [1-3]. It consists of a coherent flow of particles, which tunnel through a barrier in the presence of a chemical potential gradient. Dilute trapped Bose-Einstein condensates (BEC) represents a new system for studying Josephson effects. In particular, manipulation of optically trapped Bose-Einstein condensates, enable the investigation of dynamical regimes not easily accessible with other super-conducting or super-fluid systems. Remarkable experimental progress has led to the creation of atomic BEC Josephson Junction arrays, in which harmonically trapped atoms are additionally confined by an optical lattice potential [4,5]. In such systems, the dynamics is driven by the competition between two physical magnitudes: the Josephson coupling energy $E_J$, which governs the tunneling through the intrawell barriers, and the onsite interparticle interaction energy $E_C$. When they are comparable, there is a competition between long-range order (which is favored by $E_J$) and localization (induced by $E_C$). When the latter prevails, no net current can flow through the junctions. Phase coherence in different wells was observed by interference experiments of condensates released from the lattice [5,6]. In addition, Josephson effects [4] and the control of tunneling rate have been demonstrated [7,8].
Alternative insight into the diverse range of Josephson phenomena can be obtained by looking at a single Josephson junction arising in a double – well system. This system has already received considerable theoretical attention [9]. In this paper, we investigate the effect of dissipation, on Josephson dynamics. Such a double well could be a part of frequency chirped optical lattice. Accelerating the optical lattice leads to Bloch oscillation of the condensate in the lowest band, as well as Landau-Zener (LZ) tunneling into higher bands [10].

## Tunneling dynamics and dissipation in an accelerated double well system

As discussed in the introduction, there has been considerable interest in the phenomenon of quantum mechanical tunneling of a Bose-Einstein condensate in a double well potential. In this, content an important role is played by the question of the effect of dissipation on the tunneling process. It was found in some very early works that the effect of dissipation is always to suppress the tunneling rate [11]. In the context of a BEC, this problem has been addressed recently [12]. If such a system is described by a coordinate $q$ with which is associated a potential energy $V(q)$ that has a metastable minimum, then the system should obey in the classically allowed regime the equation of motion of the form

$$M\ddot{q}(t)+\eta\dot{q}(t)+\frac{\partial V(q)}{\partial q}=F_{ext}(t) \tag{1}$$

Here $\eta$ is the friction co-efficient and $F_{ext}(t)$ is the external force driving the system. Thus a term proportional to time rate of change of the coordinate is essential to describe dissipation. In the context of BEC, the coordinates that describe the Josephson effect are the relative phase $\phi$ between the two wells and the fractional population difference $n$ between the two wells. The main aim of the present study is to consider the tunneling of atoms between two wells with initial phase $\phi = \pi$, whose symmetry is broken by acceleration renders the Josephson coupling energy time dependent and this mechanism emerges as the source of dissipation in the present system.

Let us consider a gas of ultra cold atoms confined by two wells separated in the x-direction, and let us assume that the trapping potential is symmetric with respect to the center of the barriers: $V(-x) = V(x)$.

$$V(x) = V_0\left[\frac{\left(x - at^2\right)^2}{b^2} - 1\right]^2 \tag{2}$$

Here $a$ is the acceleration of the system. Such an accelerated double well may be a part of an accelerated optical lattice created by two counter propagating frequency chirped laser beams [10]. $2b$ is the distance between the minima of the two wells. Such a system has been studied earlier using a slightly different potential and a different approach [13].

For the description of a condensate in a double well potential, the physical parameters needed to describe the dynamics of the system are the onsite interaction $E_C$, and the tunneling energy $E_J$. These quantities are defined in a variety of ways in the literature [10]. Our calculations are based on a variational ansatz [10] of the total Hamiltonian

$$H = \frac{-\hbar^2\nabla^2}{2m} + V(x) + g\left|\psi(\vec{r},t)\right|^2 \tag{3}$$

with the interaction parameter $g = {4\pi\hbar^2 a_s}/{m}$, $a_s$ is the s-wave scattering length and $m$ the mass of a single atom. The solution of the Gross- Pitaevskii equation in this double well potential can be mapped onto a two-state model [14] by writing the wave function as a superposition of states localized in the left and right wells, $\psi_a(\vec{r})$ and $\psi_b(\vec{r})$

$$\psi(\vec{r},t) = \exp(i\phi_a)\psi_a(\vec{r},n_a) + \exp(i\phi_b)\psi(\vec{r},n_b) \tag{4}$$

where $n_a$ and $n_b$ are the number of atoms in the left and right well respectively. $\phi_a$ and $\phi_b$ are their respective phases. The wave functions $\psi_i(\vec{r},t)[i = a,b]$ can be decomposed as:

$$\psi_i(\vec{r},n_i) = \varphi_0(x)\varphi(y,z)\sqrt{n_i} \tag{5}$$

$\varphi(y,z)$ is the part of the wavefunction perpendicular to the double well.

Basing the variational ansatz for $\varphi_0(x)$ on a gaussian of the form $\varphi_0(x) = \frac{1}{\sigma^{1/2}\pi^{1/4}}\exp\left[-\frac{1}{2}\left(\frac{x}{\sigma}\right)^2\right]$, we obtain a minimum energy wave function of width $\sigma$ which expressed in units of the width $\sigma_h = \frac{2b}{\pi}\left(\frac{V_0}{E_{rec}}\right)^{-1/4}$ in the harmonic approximation of the potential wells, satisfies the condition

$$\exp\left[-\left(\frac{\sigma}{\sigma_h}\right)^2 \Big/ \sqrt{V_0/E_{rec}}\right] = \left(\frac{\sigma}{\sigma_h}\right)^{-4} \qquad (6)$$

Here $E_{rec} = \hbar^2\pi^2 / 8mb^2$ is the recoil energy. This equation can be solved numerically to yield $\sigma/\sigma_h$. The quantities $E_C$ and $E_J$ are defined as

$$E_C = N_S g_{1d} \int dx |\varphi_0(x)|^4 \qquad (7)$$

$$E_J = -\int dx \varphi_0(x)\left[-\frac{\hbar^2\nabla^2}{2m} + V(x)\right]\varphi_0(x-2b) \qquad (8)$$

$g_{1d} = g/\pi\sigma_y\sigma_z$, where $\sigma_y$ and $\sigma_y$ are the Gaussian widths in the y and z directions of the radial wave- functions. $N_S = N/2$, $N = n_a + n_b$. Figure 1 shows the dependence of the ratio $\Lambda = E_C/E_J$ as a function of time for $a = 0.10 m/\sec^2$, $V_0/E_{rec} = 0.5$, $E_C/\hbar = 2kHz$ and $\omega_{rec} = 20kHz$. Let us focus our attention on any one of the two wells. The condensate is assumed to be initially at rest in at the center of the well. As the condensate accelerates and starts to climb up the potential well, the tunneling energy increases, $\Lambda$ decreases and approaches the unphysical value of zero at a certain time, which we denote by a break down time $\tau_{breakdown}$. This shows that the two state model breaks down as the condensate approaches the top of the barrier, which according to terminology used in the literature is said to at the edge of the Brillouin zone [10,14]. As is well known, sufficiently large interactions ($E_C >> E_J$) lead to the appearance of a loop structure at the Brillouin zone edge [14]. A direct consequence of this loop structure is the breakdown of the Bloch oscillations due to nonzero adiabatic tunneling into the upper band.
The Hamiltonian for this two state model can now be conveniently rewritten as

$$H_J = \frac{1}{2}\begin{bmatrix} -\Delta\mu + E_C n & -E_J \\ -E_J & -\Delta\mu - E_C n \end{bmatrix} \tag{9}$$

Where $n = {(n_a - n_b)}/{N}$ is the fractional population difference between the left and right well, $\Delta\mu$ is the chemical potential difference between the two wells. In terms of the relative phase $\phi = \phi_a - \phi_b$, the two state schroedinger equation can be rewritten in terms of the coupled equations

$$\hbar\dot{n} = E_J\sqrt{1-n^2}\sin\phi \tag{10}$$

$$\hbar\dot{\phi} = \Delta\mu - nE_C - \frac{E_J n}{\sqrt{1-n^2}}\cos\phi \tag{11}$$

In the limit of small oscillations around equilibrium ($\phi = 0, n = 0$), the above equations (10) and (11) takes the form

$$\hbar\dot{n} = E_J\phi \tag{12}$$

$$\hbar\dot{\phi} = \Delta\mu - nE_C - nE_J \tag{13}$$

Equations (12) and (13) can be recast in two uncoupled second order differential equations (similar to eqn.(1)) in coordinates $\phi$ and $n$.

$$\ddot{n} - \frac{\dot{E}_J}{E_J}\dot{n} + \frac{E_J(E_J + E_C)}{\hbar^2}n = \frac{\Delta\mu E_J}{\hbar^2} \tag{14}$$

$$\ddot{\phi} - \frac{\dot{E}_J}{(E_J + E_C)}\dot{\phi} + \frac{E_J(E_J + E_C)}{\hbar^2}\phi = \frac{\Delta\dot{\mu}}{\hbar} \tag{15}$$

Thus we find that the dissipation term is proportional to $\dot{E}_J$. The time dependence of $E_J$ comes in through the time dependent potential. Dissipation destroys the symmetry between $\phi$ and $n$, since the rate of energy dissipation is different for $\phi$ and $n$. The effect of this dissipative mechanism on the dynamics of the system will be discussed later. Our main aim will be to solve equations (10) and (11) numerically. To find the dynamics of $\phi$ and $n$, one needs to know the values of $E_J, E_C$ and $\Delta\mu$. We will take $\Delta\mu = 0$ through out the paper. The two state model is an excellent approximation to the full solution of the schroedinger equation as long as $\Delta\mu$ and the acceleration is not to large [9]. In this simple model, the 'Bloch bands' are obtained by finding the eigenvalues of the Hamiltonian $H_J$ of equation (9)

$$H_J \begin{pmatrix} a \\ b \end{pmatrix} = \lambda \begin{pmatrix} a \\ b \end{pmatrix} \tag{16}$$

The eigenvalues $\lambda$ are also called adiabatic energy levels. Specifically for $\Delta\mu = 0$ these are:

$$\lambda = \pm\sqrt{E_J^2 + n^2 E_C^2} \tag{17}$$

If the system is initially prepared with a phase difference $\pi$, then the population tunnels back and forth and the relative phase between the two wells oscillates around a mean value of $\pi$ ($\pi$ oscillation). The amplitude of the $\pi$ oscillations depends on the ratio $\Lambda = E_C / E_J$. For $\Lambda$ less than a critical value $\Lambda_C$ (~1.28) [15], the population oscillates between $\pm n(0)$, whereas for $\Lambda > \Lambda_C$ the oscillations in $n$ are suppressed. We now proceed to study the influence of accelerating the double well on the tunneling behaviour. The effect of introducing acceleration depends sensitively on the well depth and the initial value of $\Lambda$.

In figure 2a and 2b, we have plotted the fractional relative population $n(t)$ and the eigenvalues $\lambda(t)$ as a function of time for $\Lambda(0) = 0.45$, $V_0 / E_{rec} = 0.5$, $a = 0.01 m/\sec^2$, $b = 0.2\mu m$. The corresponding break down time $\tau_{breakdown}$ is around 3 ms. Complete oscillations between the wells is observed till $\tau_{breakdown}$. During this period, the ratio $E_C / E_J$ remains close to its initial value of 0.45. As time approaches $\tau_{breakdown}$, there is a rapid oscillation in population. The time evolution of the band gap (fig 2b) shows that the energy gap (and hence $E_J$) is the lowest at $\tau_{breakdown}$. It is to be noted that all parameters beyond the break down time is unphysical and is depicted in the figures only for illustration purpose. On increasing the acceleration to 1 m/sec$^2$ (keeping other parameters the same), the population oscillations are suppressed (fig 3). The break down in this case develops much sooner ($\tau_{breakdown} \approx 0.3$ ms). Hence if the BEC is initially in a state belonging to the lowest energy band and the acceleration is small enough, the BEC will stay in the lowest energy band and undergo population oscillations between the two wells (provided $\Lambda < \Lambda_{cr}$). As one increases the acceleration the BEC will have an increasing chance of tunneling into the upper band near $\tau_{breakdown}$ (where the band-gap is smallest) and hence leading to a disruption of the population oscillation.

If the lattice depth is fairly small ($\leq 12\,E_{rec}$), then the tunneling rate calculated from Landau-Zener theory is given by

$$r = \exp\left[-\frac{\pi^2 V_0}{mda}\right] \qquad (18)$$

The effect of increasing the well depth is shown in fig 4a, where $n$ is plotted as a function of time for $V_0/E_{rec} = 5$, $a = 0.01 m/\sec^2$ (other parameters remaining same as before). At this depth, the initial value of $\Lambda$ is already greater than $\Lambda_C$ (Fig.4b). As $\Lambda$ decreases with time and approaches a value less than $\Lambda_C$, $n$ partially oscillates between the initial value of 0.5 and 1.0 till $\tau_{breakdown}$. On increasing the acceleration to $1 m/\sec^2$ there is a complete absence of population oscillation as shown in Fig.5. Let us briefly make a remark here regarding the phase. The dynamics of the phase is more or less similar to that of the population difference. Acceleration of the double well also inhibits the phase oscillation. Now let us comment on the experimental aspect of such theoretical analysis. In a recent experimental work [16], it was shown that the visibility of the interference pattern of the condensate when released from the trap improved with increase in the acceleration of the optical lattice. The loss of visibility of interference pattern was attributed to instabilities near the band edge. The visibility reflects the phase coherence of the condensate. Visibility is close to unity for a near perfect coherence. In light of the above theoretical analysis, we interpret the loss of visibility of the interference pattern at low acceleration of the double well as due to oscillations in phase $\phi(t)$ as well as $n(t)$. As the acceleration increases, the phase $\phi(t)$ as well as $n(t)$ stabilizes and this leads to restoration of visibility. At this point the role of the dissipative mechanism (due to time dependent $E_J$), which we had introduced earlier, in stabilizing the system becomes clear. Dissipation damps the population as well as phase oscillation leading to an overall stable system. The exact dynamics at large accelerations or beyond the break down time $\tau_{breakdown}$ is still not clear as the simple two-state model breaks down.

## Conclusion

We have studied the Josephson dynamics of a Bose-Einstein condensate in an accelerated double well potential. We show that acceleration can lead to suppression of population as well as phase oscillations between the two wells. The suppression of population oscillation arises because the band-gap decreases significantly and thereby increasing the probability of tunneling into the higher bands. The time dependence of $E_J$ emerges as a source of dissipation in the present system, which helps in stabilizing the system by maintaining a constant phase across the two wells. This phase coherence between the two wells is reflected as restoration of visibility of interference fringes in a recent experiment. The two-state model is found to break down when the condensate approaches the top of the well (or the edge of the band).

**References**:

1. Josephson BD 1962, Phys. Lett . **1**, 251.
2. Barone A and Paterno B 1981 ‘Physics and applications of the Josephson effect’ (New York, Willey).
3. Avenel O and Varoquank E 1985, Phys. Rev. Lett. **55**, 2704.
4. Cataliotti F S, Burger S, Fort C, Maddoloni P, Minardi F, Trombettoni A, Smerzi A and Inguscio M 2001, Science **293**, 843.
5. Orzel C, Fenslan A K, Yasuda M and Kasevich M A, 2001, Science, **291**, 2386.
6. Anderson B P and Kasevich M A 1998, Science **282**,1686.
7. Denschlag J H etal. 2002, J. Phys B. At. Mol. Opt. Phys. **35**, 3095.
8. Greiner M, Mandel O, Esslinger T, Hänsch T W and Bloch I 2002, Nature **415**, 39.
9. Sakellari E, Proukakis N P, Leadbeater M and Adam C S 2004. New J. Phys **6**, 1. See references therein.
10. Cristiani M, Morsch O, Müller J H, Ciampini D and Armondo E, 2002, PRA,, **65**, 063612, Morsch O, Müller J H, Cristiani M, Ciampini D and Arimondo E, PRL 2001, **87**, 140402.
11. Leggett, A. J. PRB 1984, **30,** 1208.
12. Dziarmaga J, Smerzi A, Zurek W. H and Bishop A R, 2002, PRL, **88**, 167001.
13. Konotop V.V, Kevrekidis P.G and Salerno M, Cond-Mat/0404608.
14. Wu B and Niu Q, New J, Phys. **5,** 2003, 104.
15. Bhattacherjee A and Mohan M, Mod. Phys. Letts. B. 2002, **16,** 1021.
16. Jona-Lasinio M, Morsh O, Cristiani M, Arimondo E and Menotti C, Cond-Mat/0501572.

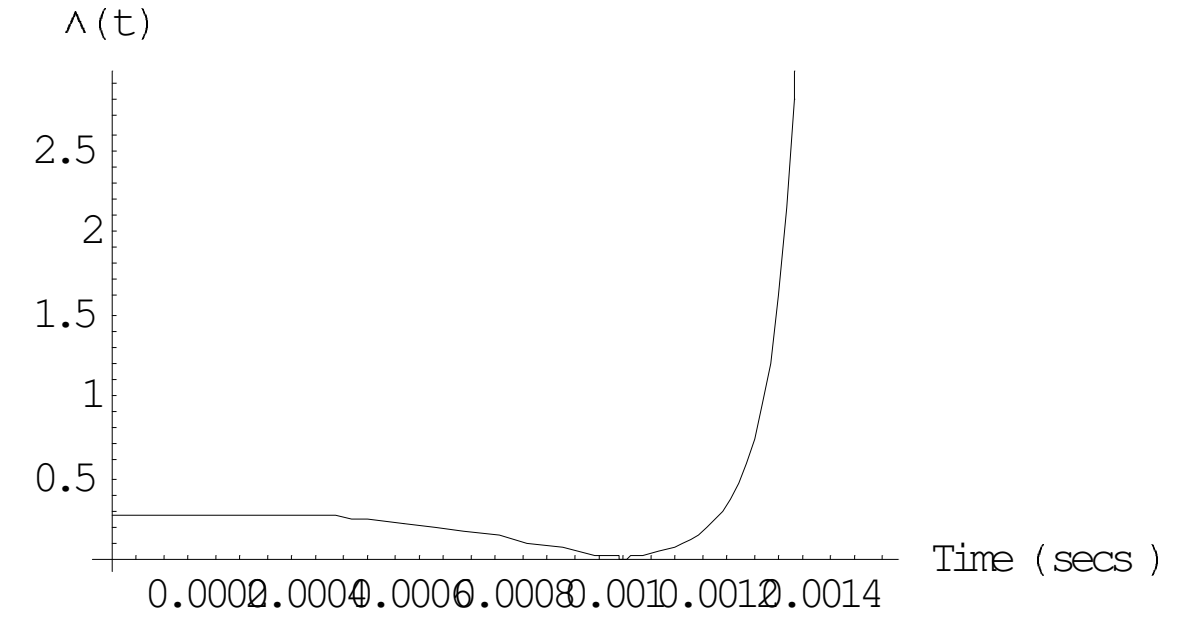

Figure1: Time evolution of the parameter $\Lambda(t) = E_C/E_J$ for $a$ =0.1 m/sec$^2$, $V_o/E_{rec} = 0.5$, $E_C/\hbar = 2kHz$ and $\omega_{rec} = 20kHz$.

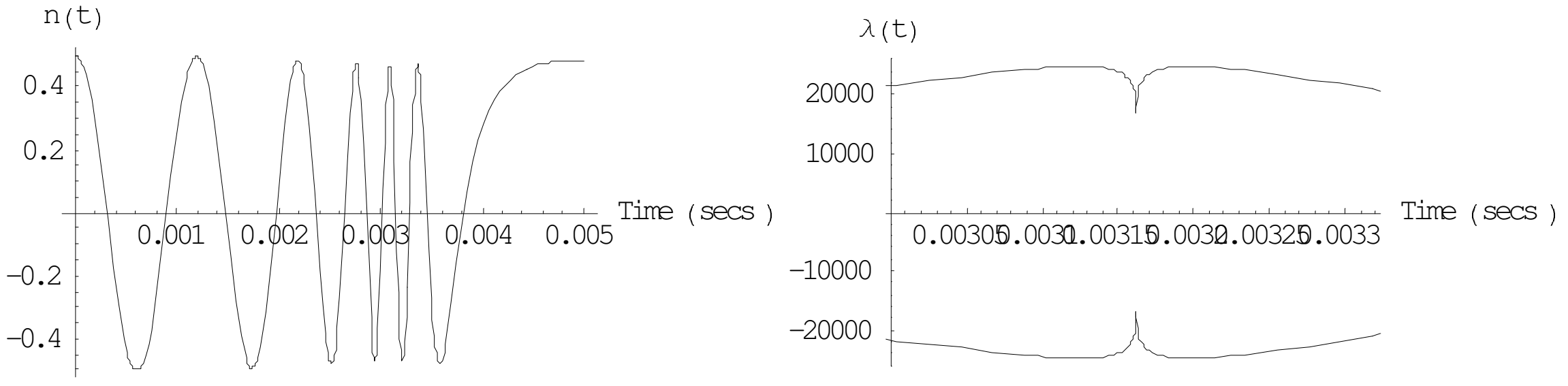

(a) (b)

Figure 2: The fractional relative population $n(t)$ and the eigenvalues $\lambda(t)$ as a function of time for $\Lambda(0) = 0.45$, $V_o / E_{rec} = 0.5$, $a = 0.01 m/\sec^2$, $b = 0.2 \mu m$.

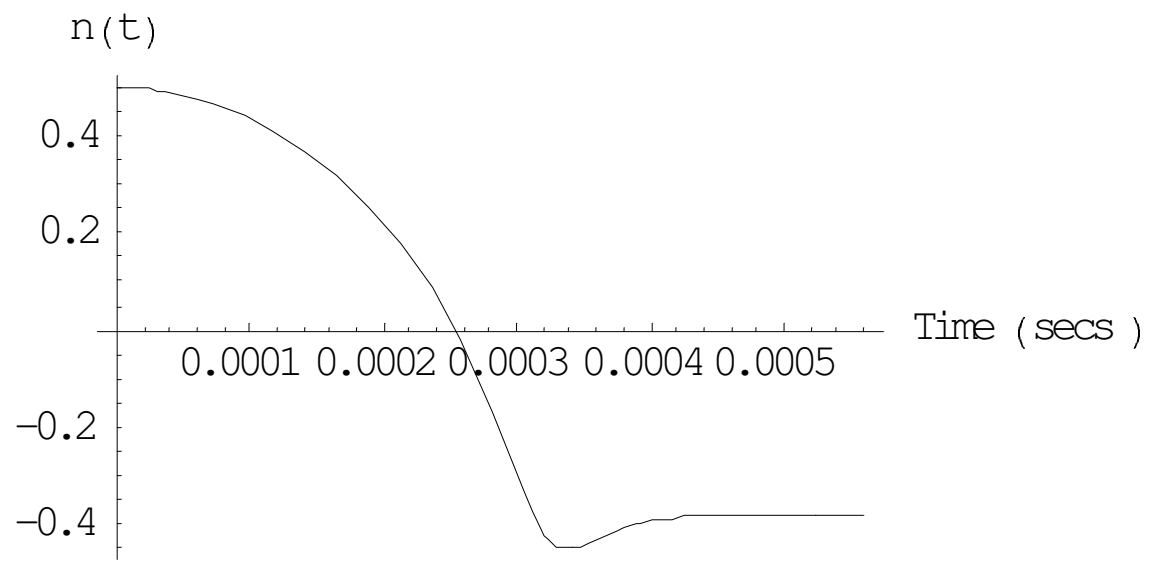

Figure3: The fractional relative population $n(t)$ as a function of time for $\Lambda(0) = 0.45$, $V_o / E_{rec} = 0.5$, $a = 1 m/\sec^2$, $b = 0.2 \mu m$.

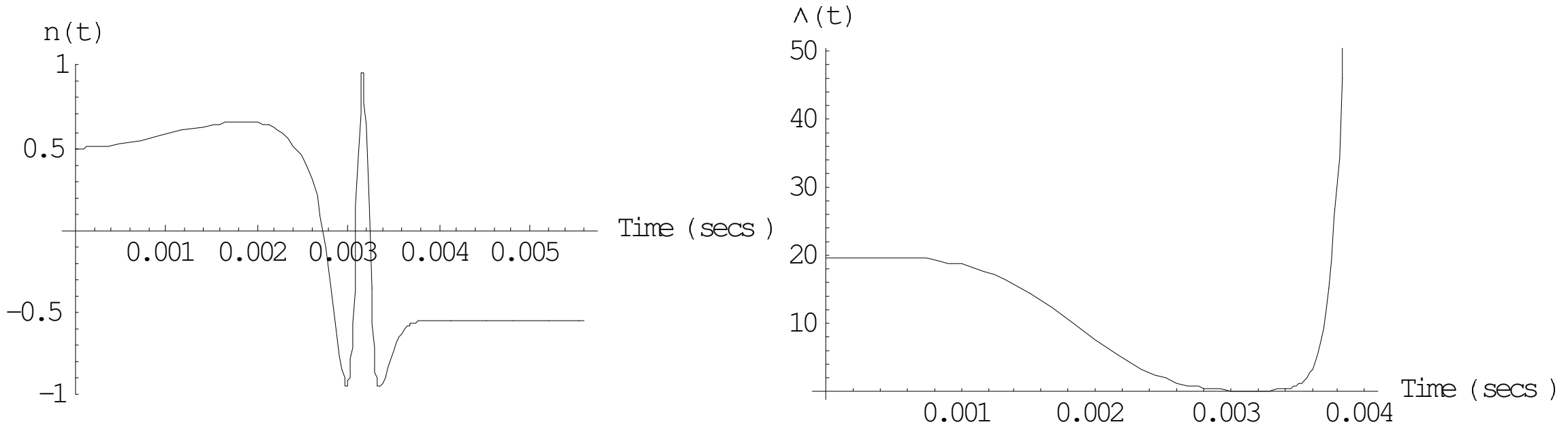

(a) (b)

Figure 4: The fractional relative population $n(t)$ and the eigenvalues $\lambda(t)$ as a function of time for $\Lambda(0) = 1.78$, $V_o / E_{rec} = 5$, $a = 0.01 m/\sec^2$, $b = 0.2 \mu m$.

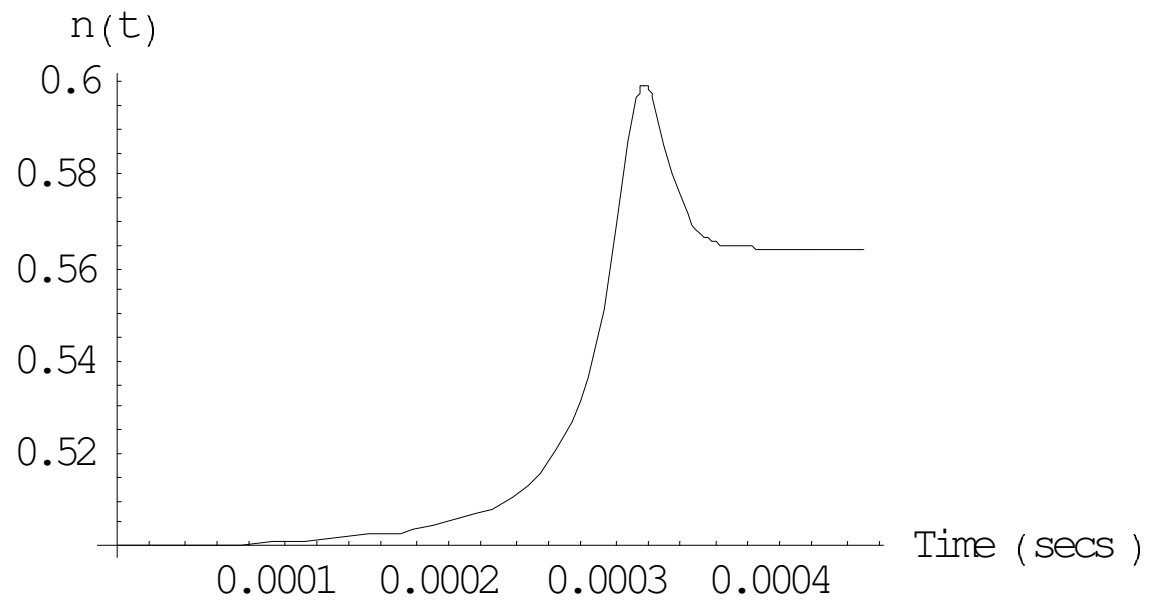

Figure 5: The fractional relative population $n(t)$ as a function of time for $\Lambda(0) = 90$, $V_o / E_{rec} = 5$, $a = 1m/\sec^2$, $b = 0.2\mu m$.